\documentclass[aps,twocolumn,pra,superscript,floatfix,superscriptaddress,showpacs]{revtex4-1}

\usepackage{epsf}
\usepackage{epsfig}
\usepackage{graphicx}
\usepackage{dcolumn}
\usepackage{braket}
\usepackage{bm}
\usepackage{amsfonts}
\usepackage{amsmath}
\usepackage{amssymb}
\usepackage{color,soul}
\usepackage{wasysym}
\usepackage{mathrsfs}
\usepackage{natbib}
\usepackage{float}
\usepackage{multirow}
\usepackage[colorlinks=true,%
            linkcolor=blue,%
            urlcolor=blue,%
            citecolor=blue,%
            filecolor=blue,%
            bookmarksopen=true,%
            pdfauthor={J. P. Ramos-Andrade},%
            pdftitle={BIC},%
            pdfsubject={BIC_Manuscript},%
            pdfpagemode=UseOutlines]{hyperref}

\newcommand{\ve}{\varepsilon}

\begin{document}

\title{Bound states in the continuum in a two-channel Fano-Anderson model}
\author{B. Grez}
\affiliation{Departamento de F\'{\i}sica, Universidad T\'{e}cnica Federico Santa Mar\'{i}a, Casilla 110 V, Valparaiso, Chile}
\author{J. P. Ramos-Andrade}
\affiliation{Departamento de F\'isica, Universidad de Antofagasta, Av. Angamos 601, Casilla 170, Antofagasta, Chile.}
\author{V. Juri\v ci\' c}
\affiliation{Departamento de F\'{\i}sica, Universidad T\'{e}cnica Federico Santa Mar\'{i}a, Casilla 110 V, Valparaiso, Chile}
\affiliation{Nordita, KTH Royal Institute of Technology and Stockholm University,
Hannes Alfvéns väg 12, SE-106 91 Stockholm, Sweden}
\author{P.\ A.\ Orellana}
\affiliation{Departamento de F\'{\i}sica, Universidad T\'{e}cnica Federico Santa Mar\'{i}a, Casilla 110 V, Valparaiso, Chile}
\

\begin{abstract}
In this article, we study the formation of the bound states in the continuum (BICs) in a two-channel Fano-Anderson model. We employ the Green's function formalism, together with the equation of motion method, to analyze the relevant observables, such as the transmission coefficient and the density of states. Most importantly, our results show that the system hosts true BICs for the case of a symmetric configuration with the degenerate impurity levels, and a complete transmission channel is then suppressed. Finally, we argue that the proposed mechanism could be relevant for the realization of BICs in the electronic and photonic systems. 

\end{abstract}

\maketitle

\section{Introduction}

The bound states in the continuum (BICs) represent exotic quantum states that coexist with a continuous band of states and, in spite of this, remain localized. As such, the BICs can be considered as zero-width resonances with infinite lifetimes coexisting with extended states. The interest in this class of states dates back to the early days of quantum mechanics, when von Neumann and Wigner predicted their formation with energies above the barrier of a particular type of spatially oscillating potential~\cite{vonNeumann-Wigner}. The most important mechanism for the formation of the BICs are the interference phenomena, which are particularly operative in electronic, photonic, and phononic systems at the nanoscale   \cite{Marinica:PhysRevLett.100.183902,azzam2021photonic,hsu2016bound,huang2020ultrafast,huang2021sound,hwang2021ultralow,Koshelev:20,pankin2020one,yu2020acousto,zhou2021geometry,cao2021elastic,Exp-PhysRevA.99.053804,Huang,melik2021fano,donarini2019coherent,Overvig-Chiral,hayran2021capturing,Longhi2021Array,Ahumada2018bound,ramos2014bound,zambrano2018bound,ladron2006fano,gonzalez2013bound,PhysRevLett.99.210404}. Furthermore, the BICs have recently attracted increasing attention, being experimentally observed in several different setups, such as an optical waveguide array structure \cite{Experimental-observation-Doelman2018,Experimental-observation-Doelman2018,PhysRevLett.111.220403}, a nonlinear photonic system through a multiphoton scattering mechanism \cite{PhysRevLett.122.073601}, patterned dielectric slab \cite{Bulgakov2017}, dielectric gratings and cylinders  \cite{Marinica:PhysRevLett.100.183902}, in electromagnetic radiation \cite{ndangali2010electromagnetic}, and in array of nanoresonators \cite{kodigala2017lasing}. They have also been discussed in the context of topological phases of matter \cite{Benalcazar-higher-order-topological,Bulgakov2017,ZhenBo2014,Takeichi2019}. On the other hand, it was demonstrated that sound confinement with an arbitrarily high-quality factor can provide a possible realization of a Friedrich-Wintgen quasi-BIC~\cite{Huang}, while a close connection between Fano resonances and quasi-BICs was recently shown experimentally in Ref.~\cite{melik2021fano}. In addition, the BICs have generated  a  great deal of attention since they provide new mechanisms to confine radiation, crucial for fundamental and technological applications. For example, the new lasers have been designed using BICs, which may be applicable in different contexts, such as optics, biological detection, and quantum information \cite{kodigala2017lasing}.

\begin{figure}[t]
\includegraphics[width=8cm]{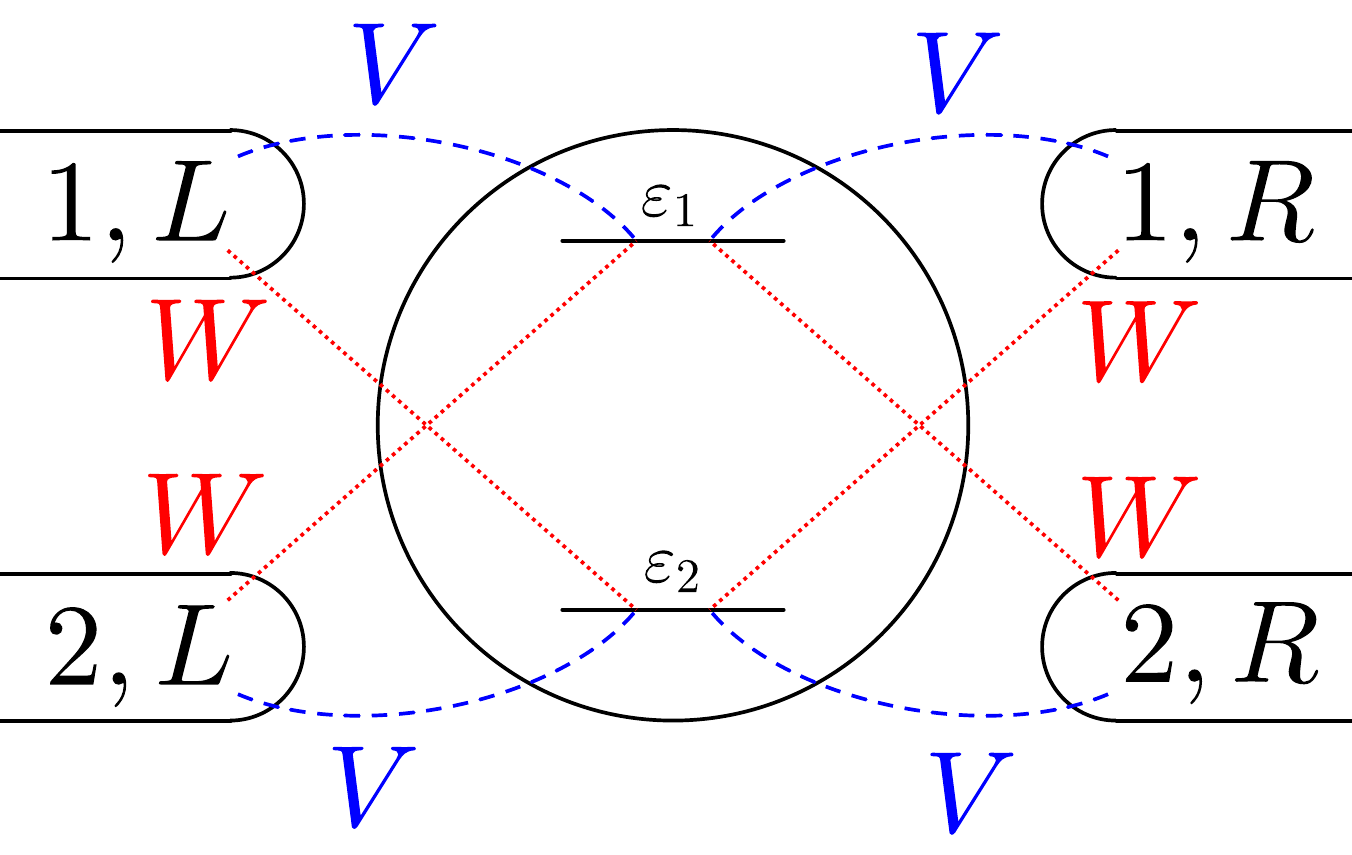}
\centering
\caption{Schematic view of the setup: a two-channel impurity coupled to left ($L$) and right ($R$) baths of excitations (bosonic or fermionic), each one featuring two degrees of freedom labeled by $1$ and $2$. The baths could be realized, for instance, as optical fibers or metallic leads, depending on the nature of the system considered. The parameter $V$ (blue dashed line) denotes the intra-channel coupling while the parameter $W$ (red dotted line) corresponds to the inter-channel coupling.}\label{fig1}
\end{figure}

In the experimental and theoretical developments mentioned above there are some common features, such as the coupling between an excitation bath and a smaller subsystem, that  eventually lead to the formation of BICs regardless of the system's details. This therefore motivates the quest for a rather simple (minimal) model featuring BICs. In this article, we provide a possible answer to this question by presenting a generic setup for obtaining BICs based on a two-channel Fano-Anderson  model. Various other incarnations of Fano-Anderson models can also host BICs~\cite{longhi2007bound,PhysRevB.80.115308}, with the mechanisms for their realization different than the one we put forward here. In this model, we include both intra- and inter-channel couplings between the left and right baths of (bosonic or fermionic) excitations, and the two-level impurity, as also illustrated in Fig.~\ref{fig1}. We describe the system analytically by employing a low-energy Hamiltonian within the Green's function formalism and using explicitly the equations of motion procedure. Our results show that the quantum interference between the two levels in the impurity,  mediated through the bath degrees of freedom (e.g. optical fiber modes or conduction electrons), produce a state strongly coupled to the baths and another one that is, in contrast, weakly coupled to them. Most importantly, in the limiting case, when the inter- and intra-channel couplings are equal, a true BIC is formed at the impurity that, as it turns out, corresponds to the antisymmetric impurity state. This manifests in the sharp peaks in the transmission coefficient (Figs.~\ref{transs} and \ref{Tcolormap}) and the density of states (Figs.~\ref{rhoo} and \ref{Fig5}). In general, when the two channels feature  different energies and are coupled, an anti-resonance is obtained due to the destructive interference, as shown in Figs.~\ref{Fig6} and ~\ref{Fig7}. We also analyze the time evolution of the impurity states using different initial conditions, as shown in Fig.~\ref{Fig8}. Finally, we briefly discuss the possible  realizations and the relevance of the BICs featured in the model in photonic and electronic systems.

The rest of the paper is organized as follows. In Sec.\ \ref{sec:modelmethod} we describe the model and outline the method employed. The results are presented in Sec.\ \ref{sec:results}. Final remarks are given in Sec.\ \ref{sec:summary}. Some technical details are relegated to Appendix~\ref{apen}. 
\newpage
\section{Model and method}
\label{sec:modelmethod}

The system is described through a two-channel Fano-Anderson  model, as it is shown schematically in Fig.\ \ref{fig1}. The corresponding low-energy Hamiltonian takes the form  

\begin{equation}\label{H}
    H=H_{\text{imp}}+H_{\text{B}}+H_{\text{imp-B}}+H_{\text{R}}\,.\\ 
\end{equation}
Here, the impurity Hamiltonian is given by
\begin{equation}\label{Himp}
    H_{\text{imp}}=\sum_\nu \ve_{\nu}d_{\nu}^\dag d_{\nu}\,,\\
\end{equation}
where $d_{\nu}^{\dag}(d_{\nu})$ is a creation (annihilation) operator corresponding to the level $\nu$ with energy $\ve_{\nu}$ ($\nu=1,2$). The Hamiltonian of the bath degrees of freedom, $H_{\text{B}}$, is given by
\begin{equation}\label{HS}
    H_{\text{B}}=\sum_{\mathbf{k}, \nu, \alpha}\ve_{\mathbf{k}, \nu, \alpha}c_{\mathbf{k}, \nu, \alpha}^\dag c_{\mathbf{k}, \nu, \alpha}\,,\\
\end{equation}
where $c_{\mathbf{k}, \nu, \alpha}^{\dag}(c_{\mathbf{k}, \nu, \alpha})$ creates (annihilates) a particle with momentum $\mathbf{k}$  and energy $\ve_{\mathbf{k}, \nu, \alpha}$ in the bath $\alpha=L,R$. The terms $H_{\text{imp-B}}$ and $H_{\text{R}}$ describe the couplings between the impurity levels and the bath quasiparticles which explicitly read

\begin{align}
    H_{\text{imp-B}}&=\sum_{\mathbf{k}, \nu, \alpha}\left(V^*c_{\mathbf{k}, \nu, \alpha}^\dag d_{\nu} +Vd_{\nu}^\dag c_{\mathbf{k}, \nu, \alpha}\right)\,,\\
    H_{\text{R}}&=\sum_{\mathbf{k}, \nu, \nu', \alpha}\left[Wd_{\nu'}^\dag c_{\mathbf{k}, \nu, \alpha}+W^*c_{\mathbf{k}, \nu, \alpha}^\dag d_{\nu'}\right]\sigma^x_{\nu\nu'}\,,
\end{align}
where $V$ and $W$ are the couplings between the $\nu$ channel of the bath $\alpha$ with the impurity channel $\nu$ and $\nu'(\neq\nu)$, respectively. 

We employ the standard  Green's function (GF) formalism to address the transport properties of the system. The elements of the retarded GF, $G^r$, obtained from the corresponding equation of motions, are given in the time domain by

\begin{equation}
    G_{i,j}^r(t)=-\frac{i}{\hbar}\Theta(t)\langle[d_i(t),d_{j}^\dag(0)]_{+(-)}\rangle\,,
\end{equation}
where $[\dots\,,\,\dots]_{+(-)}$ denotes the anticommutator(commutator). 

We focus on the transmission coefficient ($T$) across the sections and the impurity density of states ($\rho$). In terms of the GFs, these can be, respectively, expressed as 
\begin{equation}\label{tras}
T(\ve)=\text{Tr}\{ G^a(\ve)\Gamma^R G^r(\ve)\Gamma^L\}\,,
\end{equation}
and
\begin{equation}\label{densi}
\rho(\ve)=-\frac{1}{\pi}\text{Tr}\{\text{Im}\left[G^r(\ve)\right]\}\,,
\end{equation}
where $G^a(\ve)=\left[G^r(\ve)\right]^{\dag}$ is the advanced GF and $\Gamma^{\alpha}$ is the energy-independent matrix coupling between the bath $\alpha$ and the impurity. Note that Eqs.\ (\ref{tras}) and (\ref{densi}) contain the retarded GF in energy domain. The explicit form of the GF matrix elements is given in  Appendix \ref{apen},  Eqs.~\eqref{eq:GF1}-\eqref{eq:GF3}. 

Throughout this paper we use the wide band approximation to treat the intra-channel coupling, i. e. $\gamma=2\pi\rho_{0}\lvert V\lvert^2$, with $\rho_{0}$ as the bath's density of states. Within this framework, we consider a coupling between the impurity and the center of the bath’s bands that takes an approximately constant value, leading to an energy-independent $\gamma$. In turn, this  allows for a clear analytical explanation of the mechanism leading to the emergence of BICs and their features. We also set $W=\lambda V$, with $\lambda$ being a dimensionless parameter, and parametrize $\varepsilon_{1}=\varepsilon_{d}+\delta/2$, $\varepsilon_{2}=\varepsilon_{d}-\delta/2$ and $\tilde{\varepsilon}=\varepsilon-\varepsilon_{d}$, with the parameter $\delta$ as the energy difference between the impurity channels. Consequently, the transmission coefficient and the impurity density of states, respectively, are expressed as
\begin{widetext}
\begin{align}\label{eq:Tgeneral}
    T(\tilde{\varepsilon})&=8\gamma^2 \frac{4(\lambda^2-1)^4\gamma^2+(\lambda^2-1)^2\delta^2+4(1+6\lambda^2+\lambda^4)\tilde{\varepsilon}^2}{(4(\lambda^2-1)^2\gamma^2+\delta^2)^2+8(4\gamma^2(1+6\lambda^2+\lambda^4)-\delta^2)\tilde{\varepsilon}^2+16\tilde{\varepsilon}^4}\,,
\end{align}

\begin{align}\label{eq:rho-general}
    \rho(\tilde{\varepsilon})&=\frac{8(\lambda^2+1)\gamma(4(\lambda^2-1)^2\gamma^2+\delta^2+4\tilde{\varepsilon}^2)}{\pi[(4(\lambda^2-1)^2\gamma^2+\delta^2)^2+8(4\gamma^2(1+6\lambda^2+\lambda^4)-\delta^2)\tilde{\varepsilon}^2+16\tilde{\varepsilon}^4]}\,.
\end{align}
\end{widetext}

\section{Results}
\label{sec:results}

\subsection{Stationary states: Energy domain }

\begin{figure}[htb]
\includegraphics[width=0.45\textwidth]{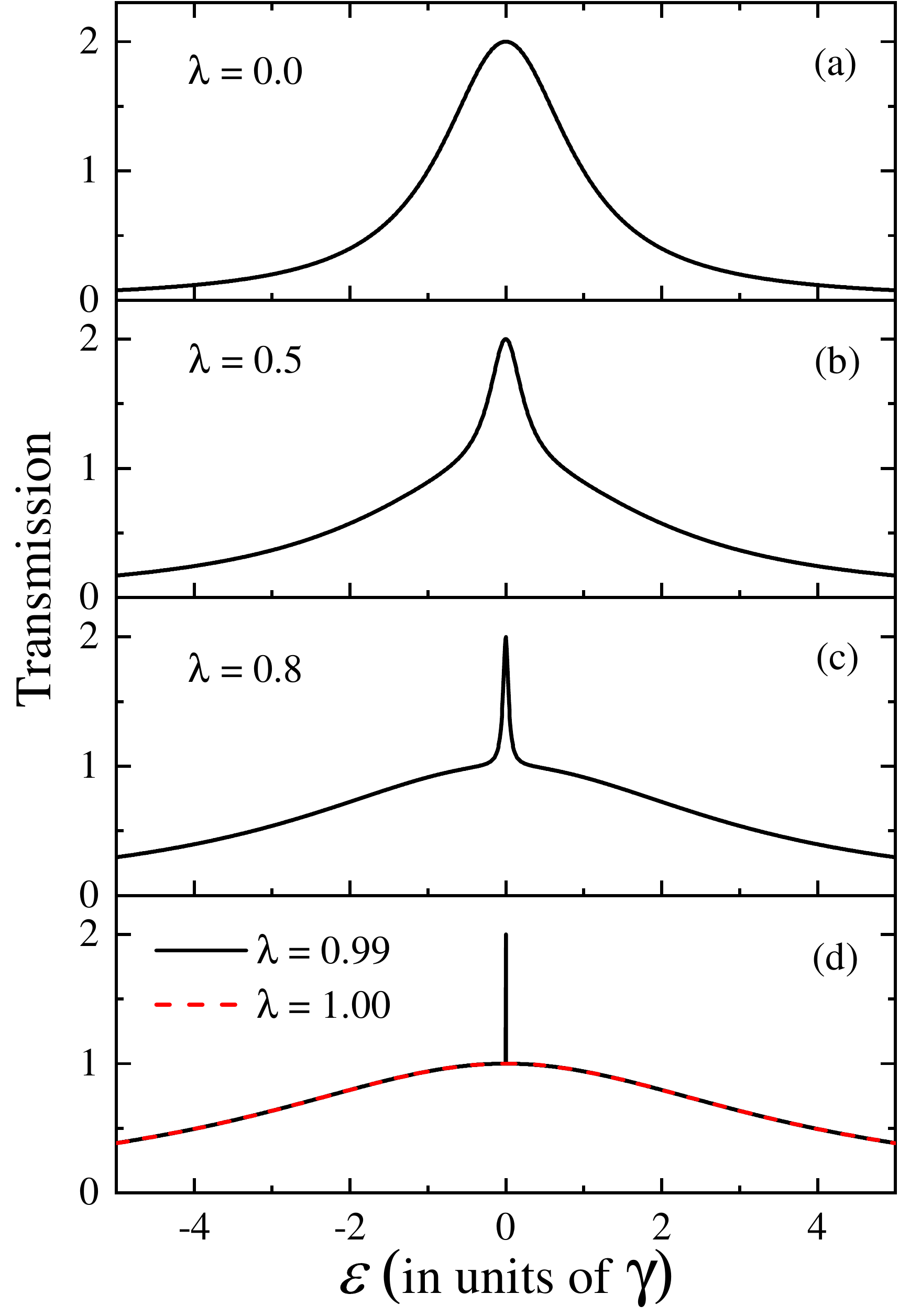}
\centering
\caption{Transmission coefficient $T$ as function of energy $\ve$ for different values of the interchannel coupling $\lambda$, as given by Eq.~\eqref{T}. (a) $\lambda=0.0$; (b) $\lambda=0.5$; (c) $\lambda=0.8$; and (d) $\lambda=0.99$ (solid black line); $\lambda=1.0$ (dashed red line). Notice that the maximum value of the transmission coefficient is equal to $2$ since two transmission channels are considered.}\label{transs}
\end{figure}

\begin{figure}[htb]
\includegraphics[width=0.45\textwidth]{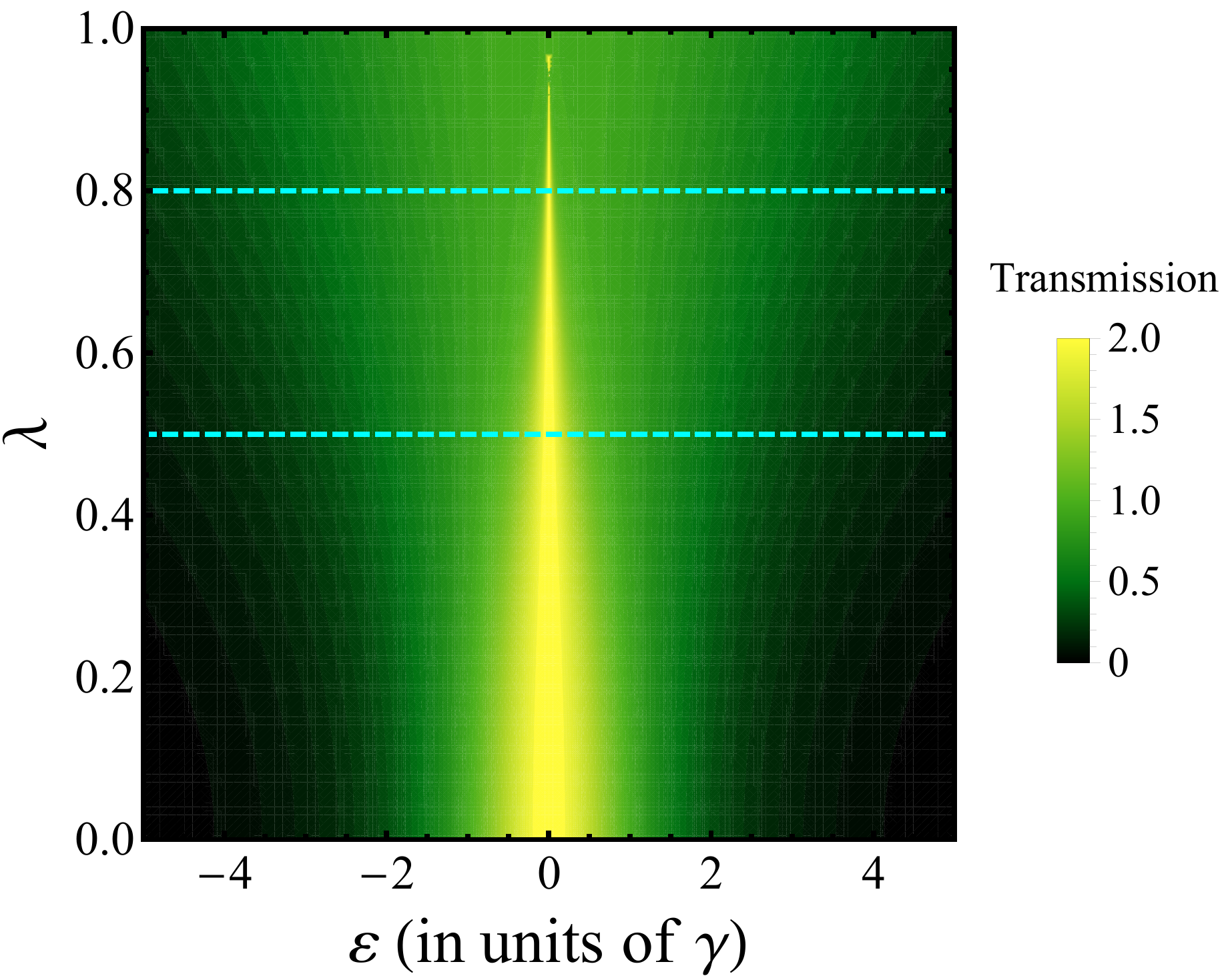}
\centering
\caption{Map of transmission coefficient $T$ as function of the energy $\ve$ and the parameter $\lambda$, which is given by Eq.~\eqref{T}. Section represented by cyan dashed horizontal line for $\lambda=0.5$ ($\lambda=0.8$) corresponds to the plot in Fig.\ \ref{transs}(b) [Fig.\ \ref{transs}(c)].}\label{Tcolormap}
\end{figure}

First we consider the case of an impurity with a symmetric spectrum by fixing $\varepsilon_{1}=\varepsilon_{2}=0$, i.e. $\ve_{d}=\delta=0$, which implies that ${\tilde \varepsilon}=\varepsilon$ in this case. We also fix the  energy scale $\gamma=1$. Figure\ \ref{transs} displays the transmission coefficient as a function of the energy of the incident particle for different values of the parameter $\lambda$. We can observe that as $\lambda$ increases the transmission coefficient evolves from the regular resonance shape [Fig.\ \ref{transs}(a)] to the superposition of both broad and thin resonances [Fig.\ \ref{transs}(b) and Fig.\ \ref{transs}(c)]. This effect  is due to the quantum interference between the paths available for the particle to cross the impurity. Interestingly, as $\lambda \rightarrow 1$, the central resonance becomes narrower, showing the broad resonance solely when $\lambda=1$. Furthermore, the color map in Fig.\ \ref{Tcolormap} displays the profile of the  transmission coefficient in terms of  energy $\ve$ and the parameter $\lambda$.  According to the above, the behavior of the transmission coefficient can be treated analytically as the superposition of two Breit-Wigner line shapes as    
\begin{equation}\label{T}
    T({\varepsilon})=\frac{\gamma^2(\lambda-1)^4}{{\varepsilon}^2+\gamma^2(\lambda-1)^4}+\frac{\gamma^2(\lambda+1)^4}{{\varepsilon}^2+\gamma^2(\lambda+1)^4}\,,
\end{equation}
where the first (second) term on the right-hand side corresponds to the thin (broad) resonance. From Eq.\ (\ref{T}) we can directly  read off the width of the broad resonance as proportional to $\gamma_{+}=\gamma(\lambda+1)^{2}$, while for the thin resonance  it is proportional to $\gamma_{-}=\gamma(\lambda-1)^{2}$.

\begin{figure}[htb]
\includegraphics[width=0.47\textwidth]{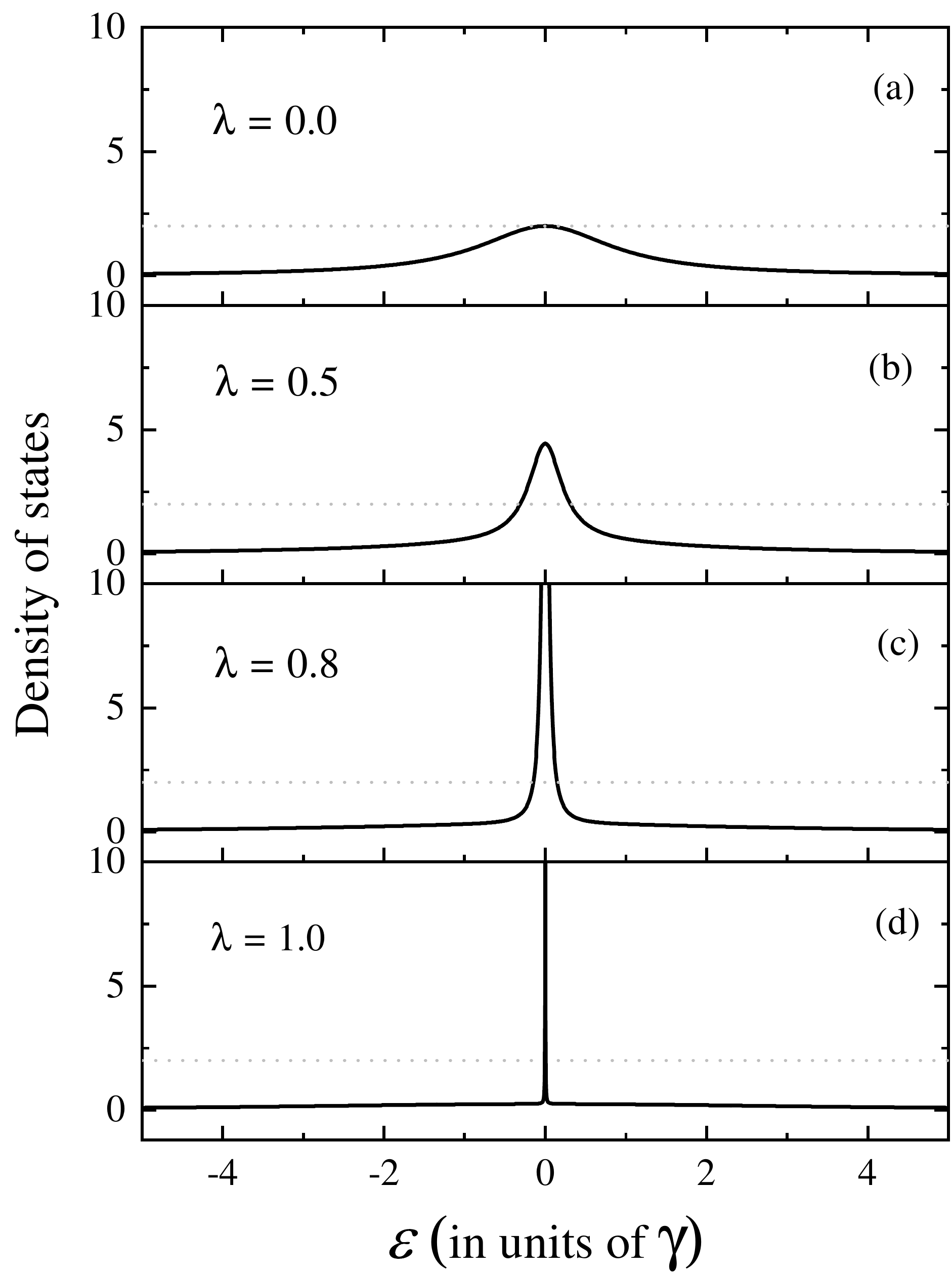}
\centering
\caption{Density of states $\rho$ as function of energy $\ve$, as given by Eq.~\eqref{rho}. (a) $\lambda=0.0$; (b) $\lambda=0.5$; (c) $\lambda=0.8$; and (d) $\lambda=1.0$. In all panels, gray dotted line represent the value $\pi\gamma\rho=2$.}\label{rhoo}
\end{figure}

The obtained  profile of the transmission coefficient is a direct consequence of the form of the impurity density of states which we analyze in the following. In Fig.\ \ref{rhoo} we show it as a function of the incident particle's energy. For the considered symmetric case, the shape of the zero energy state gradually transforms to a $\delta$-like function as $\lambda$ increases, reaching a vanishing width  when $\lambda\rightarrow 1$. Hence, we can express the density of states as
\begin{equation}\label{rho}
    \rho({\varepsilon})=\frac{1}{\pi}\left[\frac{\gamma_{-}}{{\varepsilon}^{2}+\gamma_{-}^2}+\frac{\gamma_{+}}{{\varepsilon}^{2}+\gamma_{+}^2}\right]\,.
\end{equation}

\begin{figure}[htb]
\includegraphics[width=0.45\textwidth]{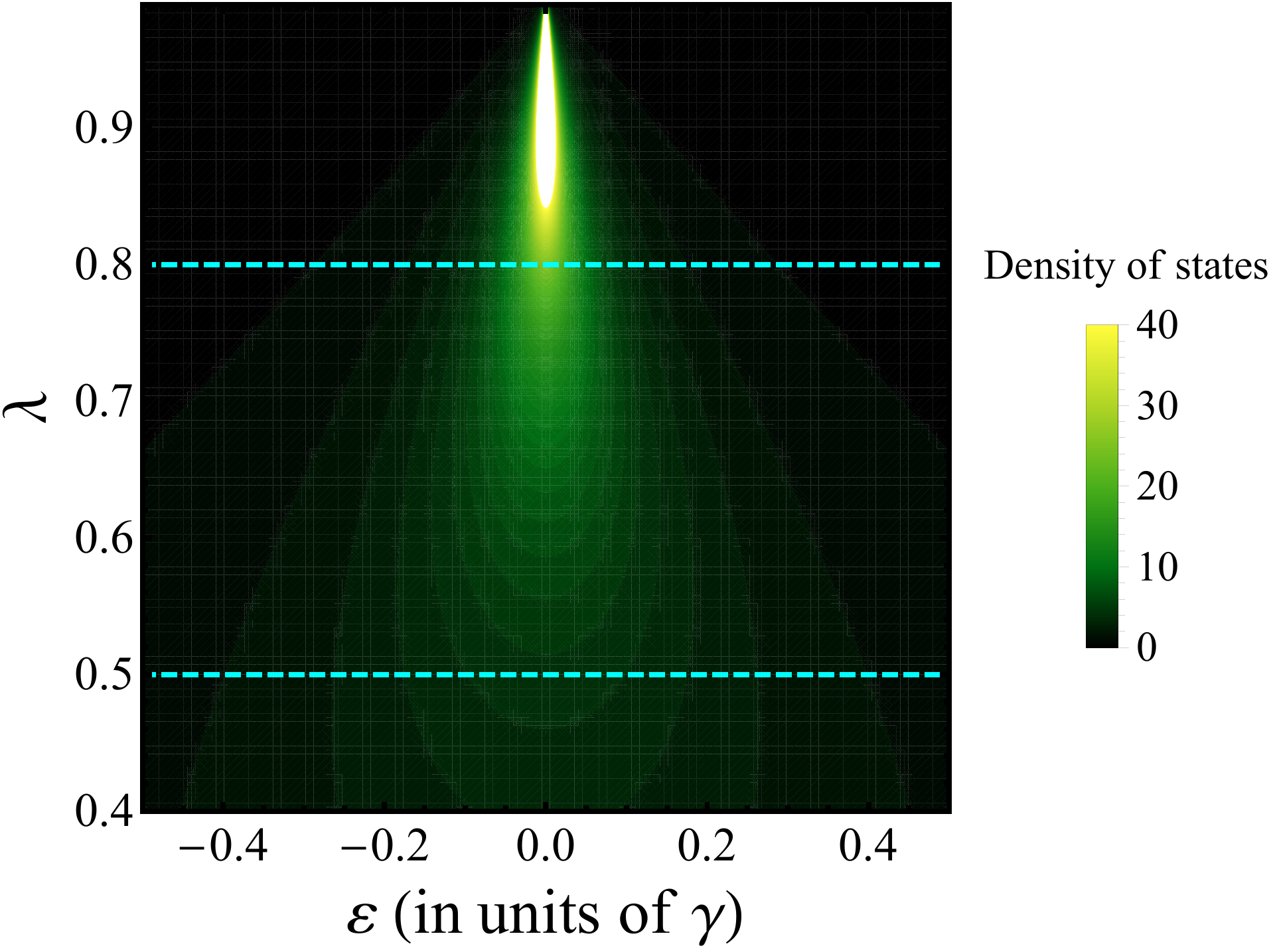}
\centering
\caption{Map of the density of states $\rho$ as a function of the energy $\ve$ and the parameter $\lambda$, which is given by Eq.~\eqref{rho}. Section depicted by cyan dashed horizontal line for $\lambda=0.5$ ($\lambda=0.8$) corresponds to the plot in  Fig.\ \ref{rhoo}(b) [Fig.\ \ref{rhoo}(c)]. The white zone observed at $\ve=0$ for $\lambda\rightarrow 1$ corresponds to higher values than the color scale used.  }
\label{Fig5}
\end{figure}

From Eq.\ (\ref{rho}), we can therefore conclude that in the limit $\lambda\rightarrow 1$, $\gamma_{-}\rightarrow 0$, the density of states reads
\begin{equation}\label{rho1}
    \rho({\varepsilon})=\delta (\varepsilon)+\frac{1}{\pi}\left[\frac{\gamma_{+}}{{\varepsilon}^{2}+\gamma_{+}^2}\right]\,.
\end{equation}
Thus, the zero-energy state becomes a BIC and the related resonance in transmission is completely suppressed. This feature can be explicitly seen in Fig.\ \ref{rhoo} as the density of states evolves from a single broad state to the superposition of both broad and $\delta$-like (BIC) states in the limit $\lambda\to1$. This behavior is also displayed  in the color map  in Fig.\ \ref{Fig5}, where the BIC corresponds to the brightest area in the plot.

\begin{figure}[htb]
\includegraphics[width=0.45\textwidth]{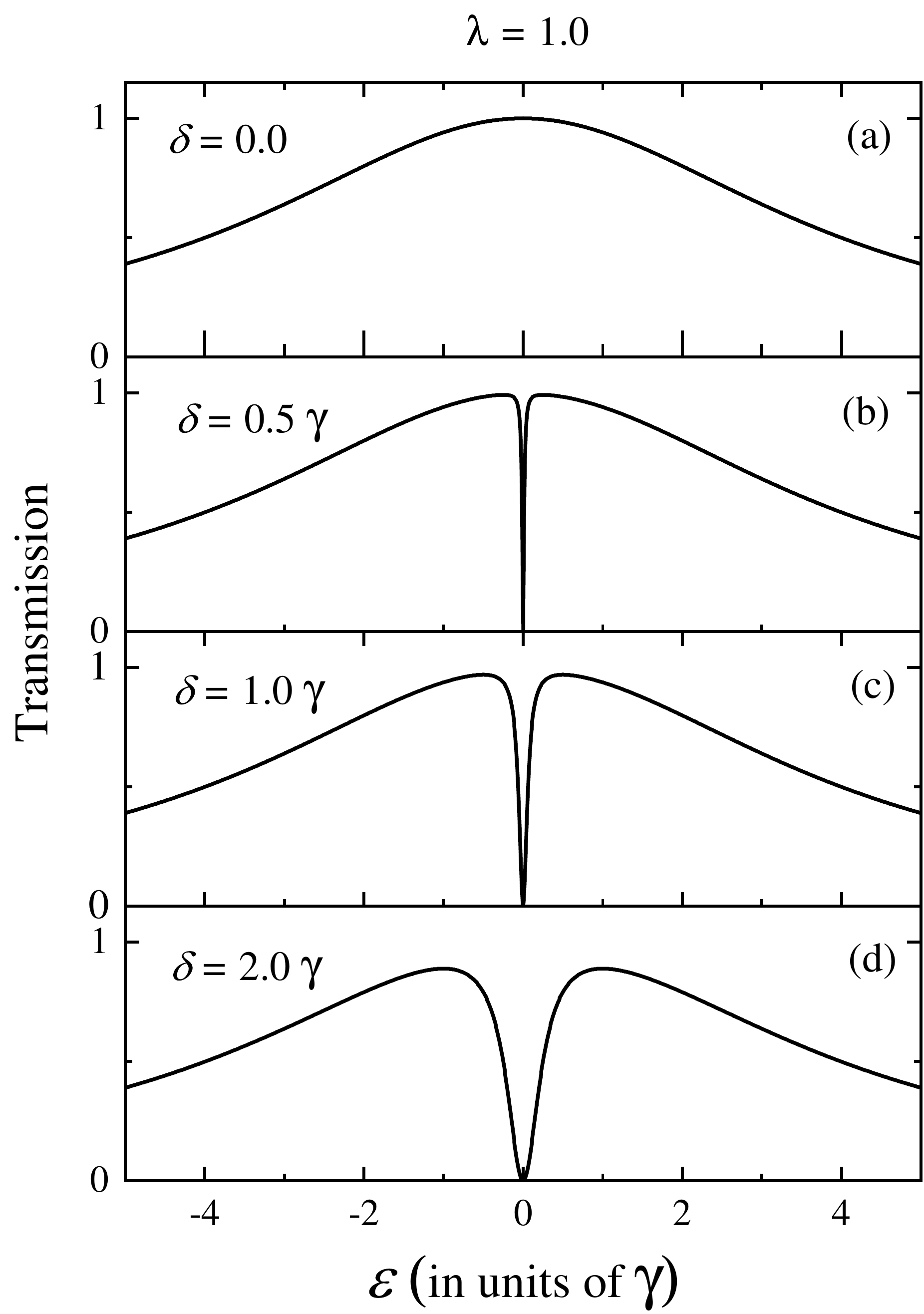}
\centering
\caption{Transmission coefficient $T$, given by Eq.~\eqref{eq:Tgeneral}, as a function of the energy $\ve$ for fixed $\lambda=1.0$ and different values of the parameter $\delta$: (a) $\delta=0.0$; (b) $\delta=0.5$; (c) $\delta=1.0$; (d) $\delta=2.0$.}
\label{Fig6}
\end{figure}

We here notice that the quasi-BIC states are always imprinted in the transmission spectrum as a Fano profile in a single-channel model. However, in the present two-channel model, the quasi-BIC state is not necessarily reflected in the transmission as a Fano-line shape because of  the degeneracy of the impurity states. Thus the different virtual paths of the particle through the two levels of the impurity are equivalent to each other. As such, the destructive interference, characteristic of the Fano effect, does not occur in this case. However, if we lift the degeneracy of the levels in the impurity by introducing a parameter $\delta$, a symmetric Fano line-shape appears in the transmission  when $\lambda=1$, as we can see in Fig.\ \ref{Fig6}.

\begin{figure}[htb]
\includegraphics[width=0.475\textwidth]{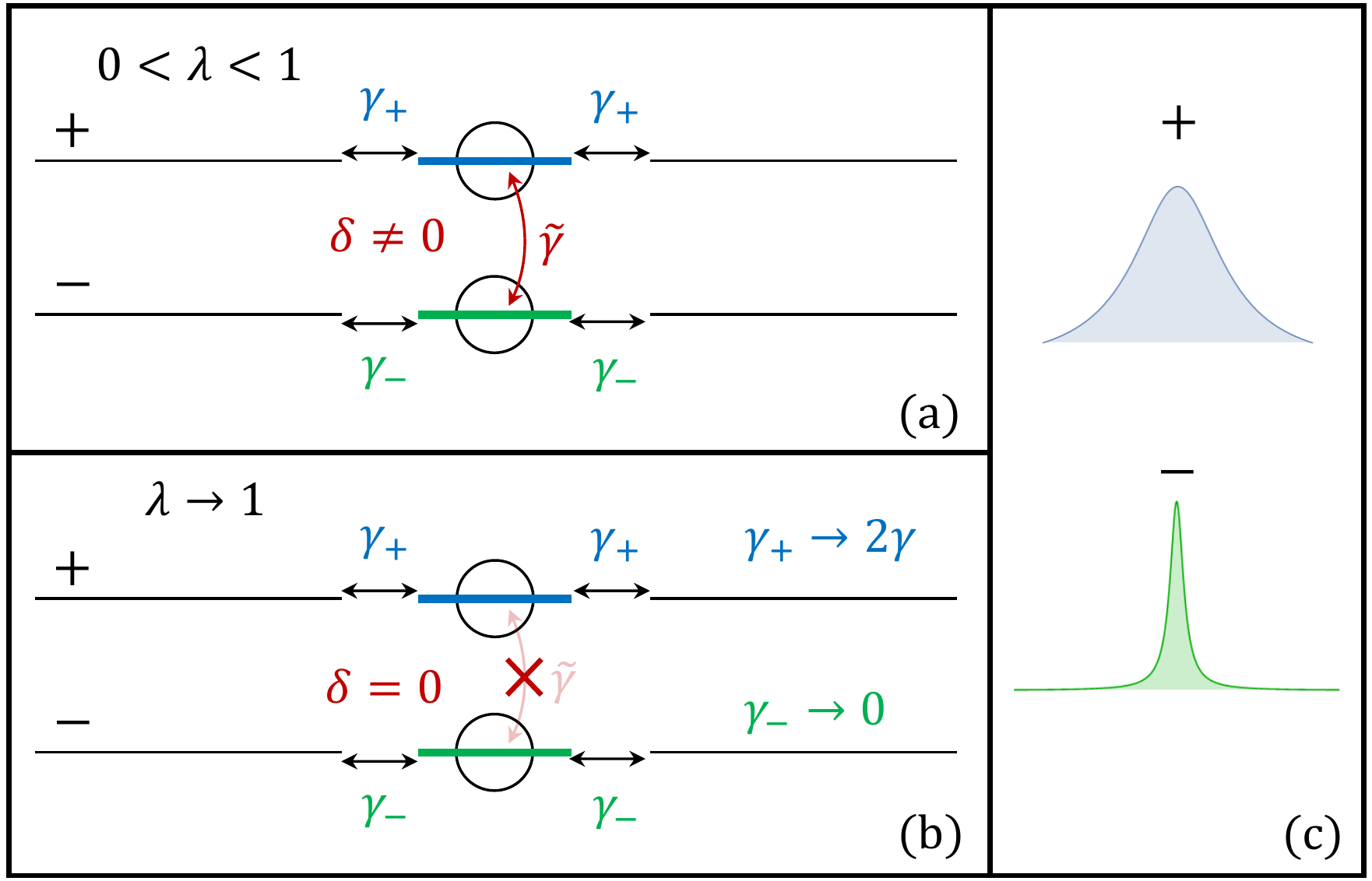}
\centering
\caption{Schematic representation of the symmetric ($+$) and anti-symmetric ($-$) channels. (a) the generic case; (b) the limiting case for $\delta=0$ and $\lambda\rightarrow 1$; (c) the shape of the resonances in the symmetric  and anti-symmetric channels.}
\label{Fig7}
\end{figure}

Our analysis so far concerns the case of the degenerate impurity states, with $\delta=0$, and  shows that the system then  behaves as two effective transmission channels, the symmetric and the antisymmetric one, labeled by $+$ and $-$, respectively. The former supports the broad resonance, while the latter hosts the sharp resonance. 
We now lift this degeneracy by considering the energy splitting $\delta\neq 0$ [see also the discussion before Eqs.~\eqref{eq:Tgeneral} and~\eqref{eq:rho-general} for the definition of the parameter $\delta$] , and moreover allows for their mixing  by means of a nonvanishing effective coupling $\tilde{\gamma}$. As it turns out, an anti-resonance then arises due to the destructive interference phenomena, as shown in Fig.\ \ref{Fig6}. In contrast, for fixed $\delta=0$ and $\lambda\rightarrow 1$,  as we have already seen, we find that $\tilde{\gamma}=0$ and $\gamma_{-}=0$, implying  that the antisymmetric channel is entirely decoupled and, most importantly, a BIC emerges. These features are summarized in Fig.\ \ref{Fig7}.  
\subsection{Time-dependent analysis}
\label{sec:tdgf}

We now move on to analyze the time evolution of the states in the impurity using the time-dependent Green's function (TDGF). 
The TDGF is given by,
\begin{equation}\label{fourier}
    G_{i,j}^{r}(t) = \frac{1}{2\pi} \int_{-\infty}^{\infty} G_{i,j}^{r}(\varepsilon) e^{-i \varepsilon t}\,\text{d}\varepsilon\,.
\end{equation}

The explicit computation of this integral is carried out in Appendix~\ref{apen}, and the final result reads
\begin{align}
    G_{1,1}^{r}(t) &= \frac{-i}{2 \Delta} \left[ \left(\Delta + \frac{\delta}{2}\right) e^{- t/ \tau_{+}} + \left(\Delta - \frac{\delta}{2}\right) e^{ t / \tau_{-}} \right]\,, 
    \nonumber \\
    G_{2,2}^{r}(t) &= \frac{-i}{2 \Delta} \left[ \left(\Delta - \frac{\delta}{2}\right) e^{- t/ \tau_{+}} + \left(\Delta + \frac{\delta}{2}\right) e^{ t/ \tau_{-}} \right]\,, 
    \nonumber \\
    G_{1,2}^{r}(t) &= G_{2,1}^{r}(t)=\frac{-\gamma \lambda  e^{-t\gamma (1+\lambda^{2})}}{\Delta} \left( e^{-it\Delta} - e^{it\Delta} \right)\,. \label{eq:tdgf12}
\end{align}
Here,  $\Delta = \sqrt{\delta^2/4 - 4 \gamma^{2} \lambda^{2}}$, and we have defined
\begin{equation}
    \frac{1}{\tau_{\pm}} = \gamma (1+\lambda^{2}) \mp i\Delta.
\end{equation}

\begin{figure}[h]
\includegraphics[width=0.475\textwidth]{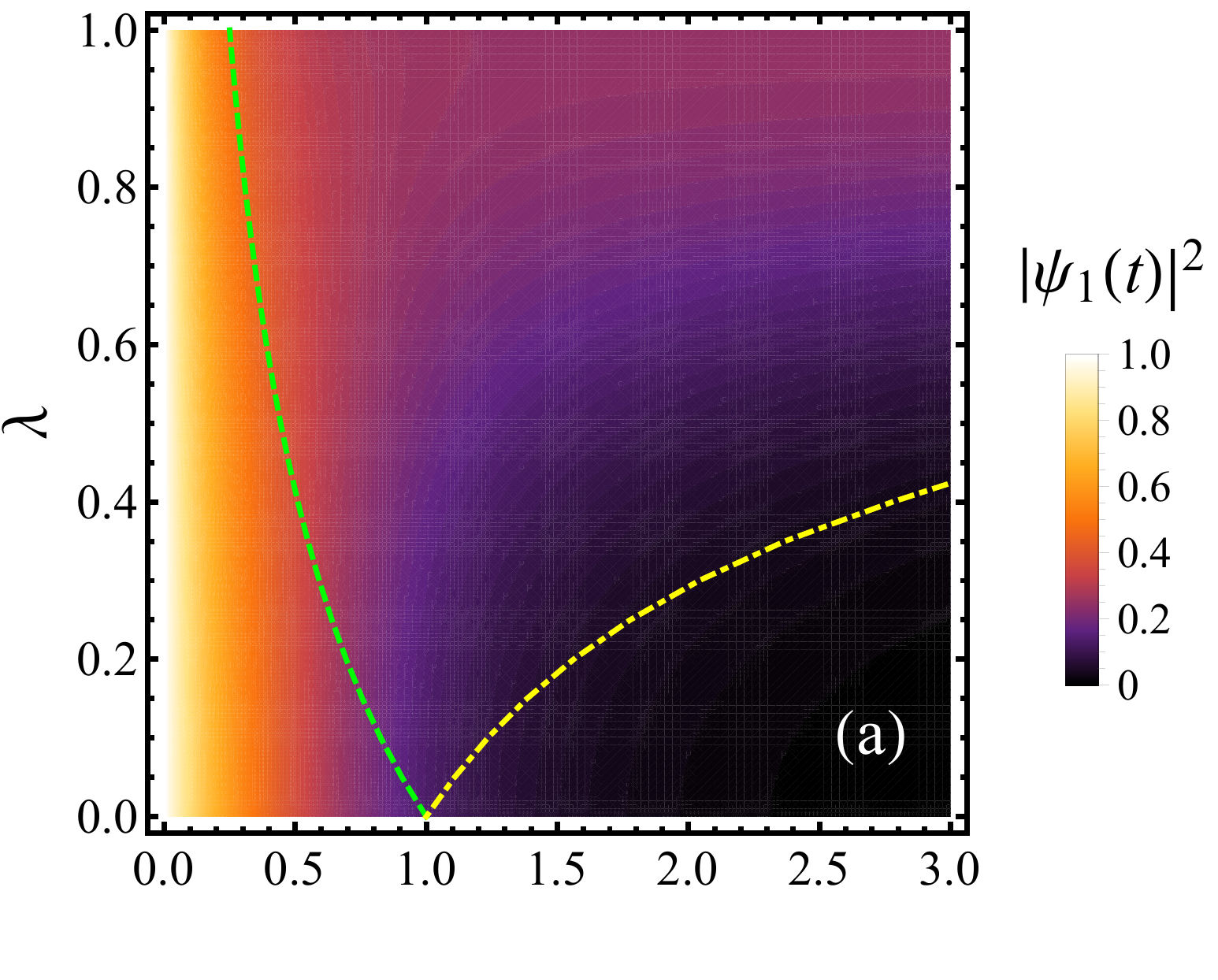}\\
\includegraphics[width=0.475\textwidth]{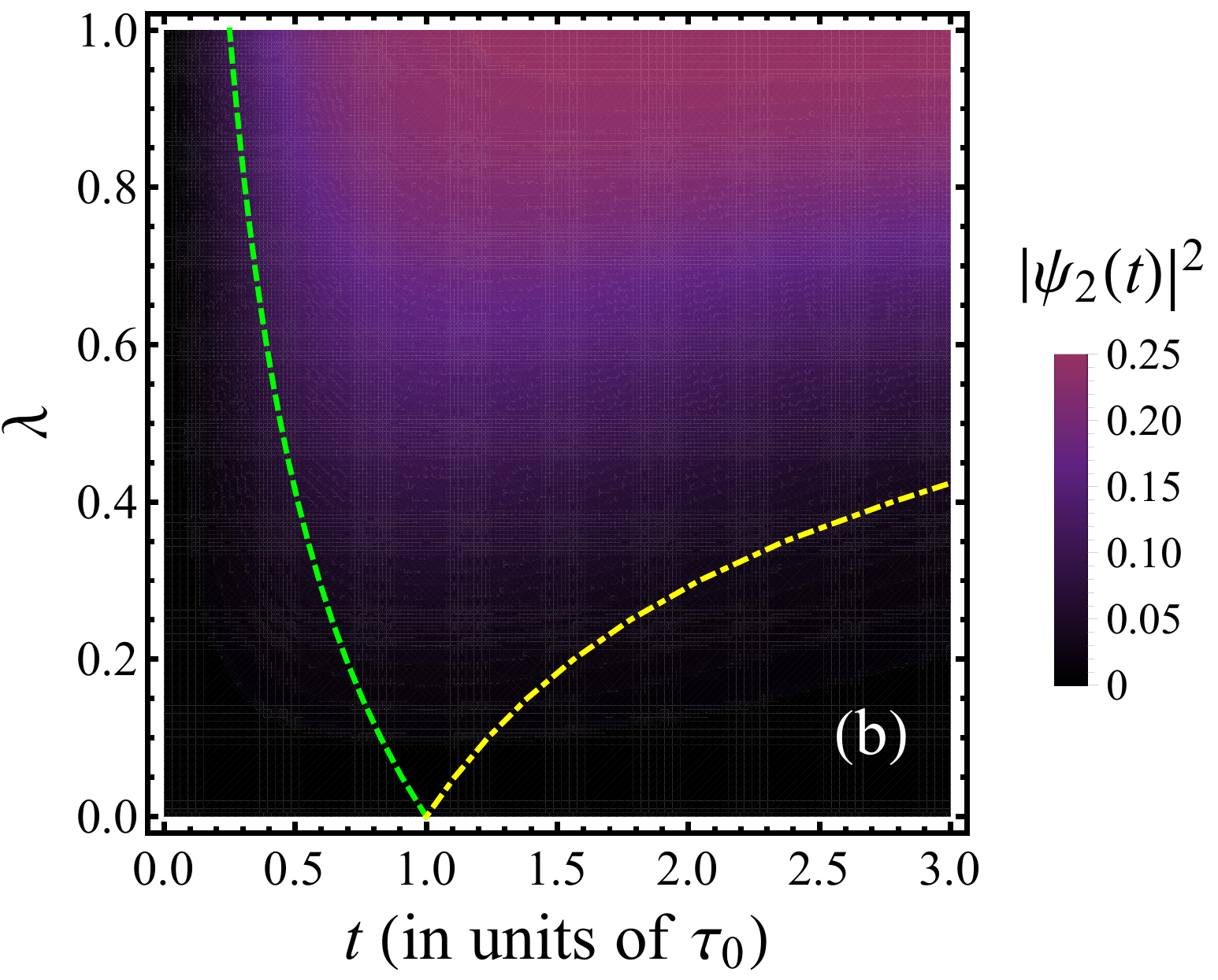}
\centering
\caption{Probability function to find the particle in the level (a) $\varepsilon_{1}$ or (b) $\varepsilon_{2}$ of the impurity given that in $t=0$ the particle is in $\varepsilon_{1}$. In both panels $\delta=0$, and the probability function is obtained from Eq.~\eqref{eq:Psi01}, together with Eqs.~\eqref{eq:tdgf12}. The green dashed and yellow dashed-dotted lines represent the values of $\tau_{+}$ and $\tau_{-}$ as a function of $\lambda$, respectively.}
\label{Fig8}
\end{figure}
If we denote the initial state by $\ket{\psi(0)}$, then the time evolution of the state will be given by
\begin{equation}
    \ket{\psi(t)}=\hat{G}^{r}(t)\ket{\psi(0)}\,,
\end{equation}
where $\hat{G}^{r}(t)$ is the retarded GF in time domain. We here use $\ket{j,\nu}$ as a basis, where $\ket{j=0,\nu}$ is the state in the impurity in the $\nu-$level. Projecting the state onto the subspace spanned by  $\ket{0,\mu}$ and inserting a complete set, we obtain the wave function of the impurity state corresponding to the level $\varepsilon_{\mu}$ ($\mu=1,2$),
\begin{equation}\label{eq:Psi01}
    \psi_{0}^{\mu}(t) = \sum_{j,\nu} G_{\mu,\nu}^{r}(t,0,j) \braket{j,\nu \lvert \psi(0)},
\end{equation}
where $\psi_{0}^{\mu}(t)=\braket{0,\mu\lvert \psi(t)}$ $G_{\mu,\nu}^{r}(t,0,j) \equiv \bra{0,\mu}\hat{G}^{r}(t)\ket{j,\nu}$ and  $G_{\mu,\nu}^{r}(t,0,0)\equiv G_{\mu,\nu}^{r}(t)$ are the functions given in Eqs.~(\ref{eq:tdgf12}). The probability of finding the particle in the level $\varepsilon_{\mu}$ of the impurity is $P_{\mu}=\lvert \psi^{\mu}_{0}(t) \lvert^{2}$.

Figure\ \ref{Fig8} displays the  probability  $P_{\mu}$ for fixed $\delta=0$ and for the following initial conditions: $\psi_{0}^{1}(0)=1$ and $\psi_{0}^{2}(0)=0$, i.e., the state 1 is fully occupied and $2$ is empty. In  Fig.\ \ref{Fig8}(a), we can clearly observe  that the time evolution of the probability is governed by two characteristic times:
\begin{equation}
\tau_{\pm}=\frac{\tau_0}{(1\pm\lambda)^2},
\end{equation}
with the time scale $\tau_0 = \hbar/\gamma$. Here,  $\tau_{+(-)}$ corresponds to the lifetime of the symmetric(antisymmectric) resonance displayed in Fig.\ \ref{Fig7}. In the limit $\lambda\rightarrow 1$, we find that $\tau_{+}=\tau_0/4$ and $\tau_{-}\rightarrow \infty$. In this case, the BIC is formed in the anti-symmetric state, and the probability first decreases quickly,  while for $t > \tau_0/2$ tends to $1/4$. The latter is expected based on the value of the  total probability being one half and the fact that a half of it leaks through the symmetric state.  In contrast, the remaining half of the total probability, which corresponds to the anti-symmetric state, is trapped at the impurity and is distributed evenly over the two available states $1$ and $2$. As the parameter $\lambda$ decreases, the anti-symmetric state then couples to the continuum of the states, and the probability decreases with time.
In contrast, for $\lambda =0$, we obtain $\tau_{+}=\tau_{-}=\tau_0$. In fact, as we can see,  for $t > \tau_0$ and $\lambda = 0$ the probability almost vanishes. For intermediate values of $\lambda$, as, for instance, $\lambda=1/2$, we find that $\tau_{+}=4\tau_0/9$ and $\tau_{-}=4\tau_0$. 
On the other hand, the probability of finding the particle in the state $2$, presented in Fig.\ \ref{Fig8}(b), remains low ($<0.05$) for rather small values of the coupling $\lambda$, such as $\lambda<0.2$. However, as the value of this parameter increases, the probability increases and tends to $1/4$ in the limit $\lambda\rightarrow 1$. Note that the sum of probabilities 1 and 2 tends to 1/2 as $\lambda\rightarrow 1$, which corresponds to the probability for the occupation of the  anti-symmetric state.  

In fact, the above behavior of the probabilities can be analytically captured by the following expressions,
\begin{align}
    P_{1}=\frac{1}{4} \left[ e^{- t/ \tau_{+}} +  e^{-t / \tau_{-}} \right]^2\,, \nonumber\\
    P_{2} =\frac{1}{4} \left[ e^{- t/ \tau_{+}} -  e^{-t / \tau_{-}} \right]^2,
\end{align}
which are obtained  using Eqs.~\eqref{eq:tdgf12} and Eq.~\eqref{eq:Psi01}. 
The total probability for finding the particle at the impurity, is then given by
\begin{equation}
    P=P_{1}+ P_{2}=\frac{1}{2} \left[ e^{- 2t/ \tau_{+}} +  e^{-2t / \tau_{-}} \right]\,. 
\end{equation}
These expressions show an exponential decay of the probabilities in the generic electronic or photonic model that we consider. We note that a non-exponential decay of a BIC was reported in a linear semi-infinite chain coupled to an impurity \cite{PhysRevA.99.010102}, which arises due to a peculiar  dependence of the surface Green's function on the energy, the form of the impurity-chain coupling, and the initial condition involving the occupation of a state orthogonal to the BIC. 

Finally, we briefly discuss the possible physical realizations of the generic model given by the Hamiltonian in Eq.~\eqref{H}. In the electronic case, the model setup can be applied to the system consisting of a single quantum dot (e. g. made of InAs) coupled with leads, with a direct diagonal coupling in spin and another one that flips the spin ($W$), which can be understood as a Rashba spin-orbit coupling. Besides, in quantum dots, where the Coulomb interaction can be significant, it is expected that the BICs may be pushed to higher energies. On the other hand, in the context of photonic systems, our model setup can be applied to the propagating photons in two waveguides that can be mixed, for instance, in a scattering region made of a whispering gallery resonator (WGR) \cite{brooks2021integrated}.

\section{Summary}
\label{sec:summary}

In this work, we investigated the formation of BICs by using a two-channel Fano-Anderson Hamiltonian. To resolve the problem, we employed both the Green's function formalism and the method based on the equations of motion. We calculated the transmission and density of states as a function of the energy for different parameters of the system, and an analysis in time-domain was performed.  Our results show that if the impurity levels are degenerate and form the symmetric configuration, the system supports BICs.   In the BIC regime, a complete transmission channel is fully suppressed. Furthermore, the time-dependent evolution of the density of states show that it is characterized by two characteristic lifetime scales, one of which is associated with the quasi-BIC state. Finally, we discuss the physical realizations of this model in the electronic and photonic platforms.

\acknowledgments
B.G. acknowledges the financial support from UTFSM master scholarship No. 034/2021. J.P.R.-A is grateful for the funding of FONDECYT Postdoc. Grant No. 3190301 (2019). V.J. acknowledges support
from the Swedish Research Council (VR 2019-04735).
P.A.O. acknowledges support from FONDECYT Grant No. 1180914 and 1201876. Nordita is partially supported by Nordforsk. 

\onecolumngrid
\appendix

\section{Time-dependent Green's function: Integral in Eq.~\eqref{fourier}} \label{apen}

In this appendix we show the steps used to compute the integral in Eq.\ (\ref{fourier}), 
which we here rewrite for completeness
\begin{equation}\label{eqapp:integral}
    G_{i,j}^{r}(t) = \frac{1}{2\pi} \int_{-\infty}^{\infty} G_{i,j}^{r}(\varepsilon) e^{-i \varepsilon t}\,\text{d}\varepsilon\,
\end{equation}
where the matrix elements of the GF read explicitly as 
\begin{align}
            &G_{1,1}^{r}(\varepsilon) = \frac{\varepsilon-\varepsilon_{2}+i\gamma(1+\lambda^2)}{K(\varepsilon)},\label{eq:GF1}\\
            &G_{2,2}^{r}(\varepsilon) = \frac{\varepsilon-\varepsilon_{2}+i\gamma(1+\lambda^2)}{K(\varepsilon)},\label{eq:GF2}\\
            &G_{1,2}^{r}(\varepsilon)=G_{2,1}^{r}(\varepsilon)=\frac{-2i\gamma h}{K(\varepsilon)}.\label{eq:GF3}
\end{align}
We here defined the function
\begin{equation}\label{eq:K}
        K(\varepsilon)=[\varepsilon-\varepsilon_{1}+i\gamma(1+\lambda^2)][\varepsilon-\varepsilon_{2}+i\gamma(1+\lambda^2)]+4\gamma^{2} \lambda^{2}.
\end{equation}

\begin{figure}[htb]
\includegraphics[width=0.475\textwidth]{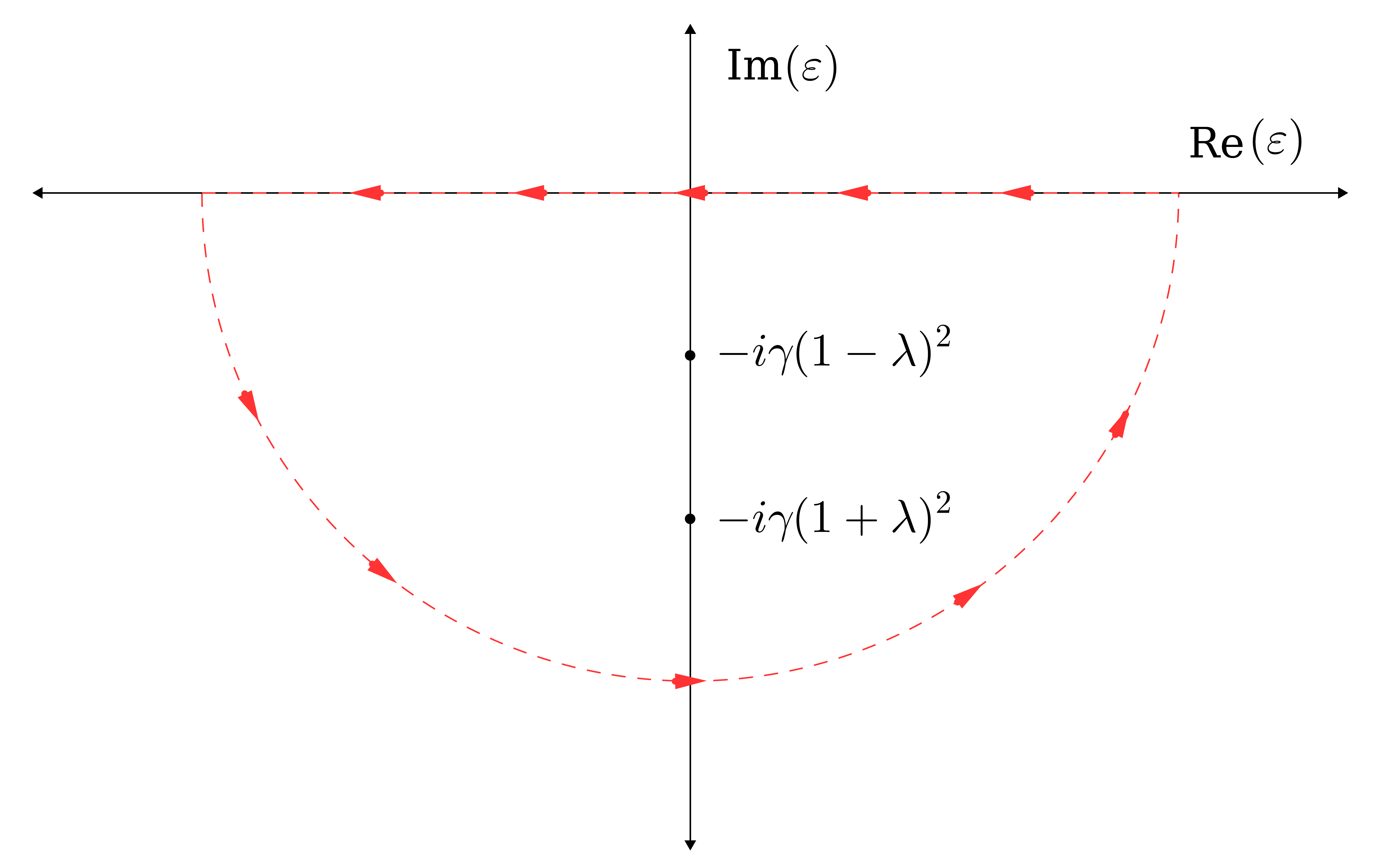}
\centering
\caption{Integration contour in the $\varepsilon$-complex plane used to compute the integral in Eq.~\eqref{eqapp:integral}}. 
\label{Fig9}
\end{figure}

We use the Cauchy residue theorem with the contour displayed in  Fig.\ \ref{Fig9}.  The two poles lying on the imaginary axis are enclosed by the red dashed line contour oriented in the  anticlockwise direction. The poles of both GFs in the energy domain, $G_{1,1}^{r}(\varepsilon)$ and $G_{2,2}^{r}(\varepsilon)$, given in Eqs.~\eqref{eq:GF1} and \eqref{eq:GF2}, in  the $\varepsilon$-complex plane can be found from the function $ K(\varepsilon)$ given by Eq.~\eqref{eq:K}, and using that $\varepsilon_{1,2}=\pm\delta/2$ in the case of the two impurity states symmetrically split about zero energy,
\begin{equation}\label{poles}
   [\varepsilon - \delta/2 + i\gamma(1+\lambda^2)][\varepsilon + \delta/2 + i\gamma(1+\lambda^2)] + 4\gamma^{2} h^{2} = 0\,.
\end{equation}
Consequently, this quadratic equation gives the form of the poles
\begin{equation}\label{eq:epsilonplusminus}
\varepsilon_{\pm} = -i\gamma (1+\lambda^{2}) \mp \sqrt{\delta^2 /4 -4\gamma^2 \lambda^2}\,.
\end{equation}
As it turns out, they are always located under the real axis (regardless of the sign of the expression under the square root), which is a necessary condition for $G_{i,i}^{r}(\varepsilon)$ to be a retarded GF. For $t>0$, the integration must be carried out in the lower half-plane (see  Fig.\ \ref{Fig9}). Then the summation of the residues at the poles given above  yields the result shown in Eqs.~\eqref{eq:tdgf12} in the main text, for the symmetric case with $\delta=0$ .

\twocolumngrid

\bibliographystyle{apsrev4-1}
\bibliography{biblio}

\end{document}